\documentstyle[aps,epsfig]{revtex}

\begin{document}



\vspace{2.5cm}
\begin{center}
   {\Large \bf Power Law in Hadron Production   
   }  
\end{center}

\vspace{0.5cm}

\begin{center}

{\bf Marek Ga\'zdzicki}\footnote{E--mail: Marek.Gazdzicki@cern.ch}\\
\vspace{0.3cm}
CERN, Geneva, Switzerland \\
and\\
Institut f\"ur Kernphysik, Universit\"at Frankfurt\\
D--60486 Frankfurt, Germany\\[0.8cm]

{\bf Mark I. Gorenstein}\footnote{E--mail: goren@th.physik.uni-frankfurt.de}\\
\vspace{0.3cm}
Institut f\"ur Theoretische Physik, Universit\"at  Frankfurt\\
D--60054 Frankfurt, Germany\\
and\\
 Bogolyubov Institute for
Theoretical Physics,
Kiev, Ukraine

\end{center}

\vspace{0.5cm}

\begin{abstract}
\noindent
In high energy p($\overline{\rm{p}}$)+p 
interactions the mean multiplicity and transverse mass spectra of
neutral mesons from $\eta$ to $\Upsilon$
($m \cong 0.5\div 10$~GeV/c$^2$) 
and the transverse mass spectra of  pions
($m_T >$ 1~GeV/c$^2$) reveal a remarkable 
behaviour: they follow, over more than 10 orders of magnitude, 
the power--law function: 
$C\cdot m_{(T)}^{-P}$.
The parameters $C$ and $P$ are energy dependent, but similar
for all mesons produced at the same collision energy.
This scaling resembles that expected in the statistical 
description of hadron production: the parameter $P$ plays the role of
a temperature and the normalisation constant 
$C$ is analogous to  the system volume.
The fundamental difference is, however, in the form of the 
distribution
function. 
In order to reproduce the experimental results and preserve
the basic structure of the statistical approach
the Boltzmann factor $e^{-E^*/T}$ appearing in standard
statistical mechanics has to be substituted by a power--law factor
$(E^*/\Lambda)^{-P}$.

\end{abstract}


\newpage

It is well established that basic properties of hadron
production in high energy collisions 
approximately follow simple
rules of statistical mechanics.
A principal assumption of
statistical models of strong interactions 
\cite{Ha:65} states that
the  single particle energy distribution in the local rest frame
of hadronizing matter follows a Boltzmann form:
\renewcommand{\theequation}{B.1}
\begin{equation}\label{exp}
\frac{dn}{d {\bf p}^*}~\sim~ \exp\left(-\frac{E^*}{T}\right)~,
\end{equation}
where $E^* = \sqrt{m^2 + {\bf p}^{*2}}= m_T \cdot {\rm cosh}(y^*)$ is 
the hadron energy 
($m_T \equiv \sqrt{m^2 + p_T^2}$ is the transverse mass and $y^*$
is the local hadron rapidity)
and $T$
is a common (for all particles) temperature parameter 
usually extracted by a  
comparison with the experimental data.
In the fixed rest frame (e.g. in the c.m. frame of
the colliding particles) the energy $E^*$ equals to
$m_T \cdot {\rm cosh}(y-y^{\prime})$, where $y$ and $y^{\prime}$
are the rapidities of considered particle and the `hadron fluid
elements' (fireballs), respectively.

Integration of hadron distribution   
along the collision axis (i.e. over $y$ and $y^{\prime}$
with an arbitrary rapidity distribution function, $f(y^{\prime})$,
of fireballs) leads to
an approximately (for $m_T >> T$) exponential form of 
transverse mass spectrum \cite{Ha:65}: 
\renewcommand{\theequation}{B.2}
\begin{equation}\label{mTexp}
\frac{dN}{m_Tdm_T}~ \sim ~\exp\left(-~\frac{m_T}{T}\right)~.
\end{equation}
According to this approach the final state hadrons should obey
this distribution providing that collective transverse 
motion of hadronizing matter and contributions from resonance decays
are neglected.
Further integration over transverse mass yields an expression
for the mean hadron multiplicity, $N(m)$, which is also (for $m>>T$) governed
by the Boltzmann factor:
\renewcommand{\theequation}{B.3}
\begin{equation}\label{mexp}
N(m)~ \sim~ \exp\left(-~\frac{m}{T}\right)~.
\end{equation}
The exponential distributions (\ref{mTexp},\ref{mexp})
are confirmed in high energy particle collisions by numerous experimental
results on $p_T$ spectra in the transverse
momentum region $p_T \leq 2$ GeV/c
and on hadron yields
at $m \leq 2$ GeV/c$^2$ \cite{Be:97}
(a correction for contributions from resonance decays is necessary
for a detailed comparison with data). 
For a given hadron species, $h$, the normalisation factors
in Eqs.(\ref{mTexp},~\ref{mexp}) are proportional to
a volume parameter $V$ (a sum of the proper volume of
`hadron fluid elements'), which has to be common for all hadrons,
following the assumption of statistical
production.
Additional factors are degeneracy factor $g = (2j+1)$, where $j$
is the particle spin, 
and a chemical factor $\exp(\mu_h/T)$, which accounts for
material conservation laws in grand canonical approximation.
The chemical potential
$\mu_h$ equals to zero for neutral hadrons (i.e. for hadrons
with zero values of the conserved charges).
Consequently, for neutral hadrons the exponential distributions
(\ref{mTexp},~\ref{mexp}) 
with a common temperature parameter and a common 
volume parameter (multiplied by the degeneracy factors $g$) are
expected to be valid.

\vspace{0.3cm}
However, in high energy p+p interactions the measurements 
of $\pi^0$ meson spectra in a high $p_T$ ($p_T > 2$~GeV/c) domain
show a strong deviation from the exponential shape.
In this region the distribution follows a power--law dependence
\cite{Da:80}. 
In the standard, QCD--based, approach
this behaviour is attributed to the hard parton
scattering \cite{parton}.
A strong violation of the Boltzmann behaviour 
in the high mass region
is also
seen in the case of 
mean hadron multiplicities.
The hadron yields
 from the $\eta$-meson ($m \cong$~0.55~GeV/c$^2$)
to the $J/\psi$-meson ($m \cong$~3.1~GeV/c$^2$) follow \cite{Ga:99}
approximately  an exponential law (\ref{mexp}).
This is, however, not true any more for
the $\Upsilon$ meson ($m \cong$~9.5~GeV/c$^2$).
The experimental ratio of $\Upsilon$ to 
$J/\psi$ mean multiplicity at 800 GeV/c  is \cite{Al:96}
$$
\frac {\langle \Upsilon \rangle} {\langle J/\psi \rangle}~
\approx~ 10^{-3}~,
$$
whereas Boltzmann's exponential law predicts:
$$
\frac {\langle \Upsilon \rangle} {\langle J/\psi \rangle}~
\approx ~\exp\left(- \frac{m_{\Upsilon} - m_{J/\psi}}{T}\right)~
\approx ~10^{-16}~
$$
for typical value of $T \cong$~0.170 GeV.
Thus the standard statistical model underpredicts the 
experimental data  by more than 10 orders
of magnitude!

\vspace{0.3cm}

In this letter we discuss a possibility to extend the statistical
approach of hadron production in elementary hadronic
interactions to the high 
(transverse) mass domain. 
Based on  experimental data we postulate 
that the energy spectrum in the local rest frame
of hadronizing matter is given by a 
power--law distribution:
\renewcommand{\theequation}{P.1}
\begin{equation}\label{power}
\frac{dn}{d {\bf p}^*} ~\sim~ \left(\frac {E^*} {\Lambda}\right)^{-P}~,
\end{equation}
instead of the exponential Boltzmann distribution (\ref{exp}).
We assume further that the remaining structure of the statistical
approach is unchanged.
Two new parameters appear in the proposed 
statistical power--law model:
a scale parameter $\Lambda$
and an exponent $P$, both are assumed to be common for all
hadrons.
We do not attempt here to introduce 
proper canonical treatment of material conservation laws
needed for  description of charged hadrons in
small systems
\cite{Ra:80,Be:97}.
Therefore, we limit our analysis to the case of
neutral mesons only.
Further on, in order to  be able to neglect  effects of
large resonance widths in the  statistical treatment
we consider only narrow mesonic states.  

Integration of Eq. (\ref{power}) over longitudinal motion 
results in a power--law transverse mass distribution:
\renewcommand{\theequation}{P.2}
\begin{equation}\label{mTp}
\rho(m_T)~\equiv ~ 
\frac {dN} {g~m_T^2~dm_T} ~= ~
C \cdot  m_T^{-P}~,
\end{equation}
where $C$ is a normalisation parameter in which we absorbed
the dependence on
the scale ($\Lambda$) and volume ($V$) parameters.
As both $\Lambda$ and $V$ are assumed to be common
for all hadrons, so does the normalisation
parameter $C$.

The integration of Eq. (\ref{mTp}) over  transverse mass yields:
\renewcommand{\theequation}{P.3}
\begin{equation}\label{mp}
\rho(m)~\equiv ~ \frac {P-3}{g~m^3} \cdot N(m)~ =
 ~C \cdot m^{-P}~,
\end{equation}
where $N(m)$ is the hadron multiplicity.

Several nontrivial predictions of the  
statistical power--law model follow from 
Eqs. (\ref{mTp}) and (\ref{mp}):
\begin{itemize}
\item
both transverse mass spectra and hadron yields should
obey a power--law behaviour;
the power $P$ in the $m_T$--distribution, $\rho(m_T)$
(\ref{mTp}), and in the hadron yield formula, $\rho(m)$ (\ref{mp}),
 should be the same,
\item
the normalisation constant $C$ also should be the same 
for both  $m_T$--distribution
(\ref{mTp}) and hadron yield spectrum (\ref{mp}),
\item
the power $P$ and the normalisation constant $C$ should be universal
(equivalent to the temperature and volume parameters in the standard
statistical approach),
i.e. they are the same for different hadron species.
\end{itemize}

In order to test these predictions we plot in Fig.~1
experimental results on $m_T$--spectra and hadron yields for
p+p interactions at $\sqrt{s} = 27\div30.8$~GeV 
as a function of $m_T$ and $m$, respectively.
Within the statistical power--law
model,
results on both $\rho(m)$ vs $m$ and $\rho(m_T)$  vs $m_T$  should 
follow the same dependence: $C\cdot m_{(T)}^{-P}$.

Full dots in Fig.~1 indicate results on $\rho(m)$ calculated  
for the mean multiplicity of hadrons: $\eta, \omega, \phi~
\cite{Ag:91}, J/\psi,
\psi'$ and $\Upsilon$ \cite{Al:96}, whereas  triangles indicate the
data on $m_T$--spectra, $\rho(m_T)$, of neutral pions \cite{pi30}.
The measurements of $J/\psi, \psi'$ and $\Upsilon$ \cite{Al:96}
were performed for p+S interactions at 800 GeV/c.
The extrapolation to p+N interactions was done assuming
the $A^{0.92}$ dependence on nuclear mass number.
The measured energy dependence of midrapidity yield \cite{Ko:80}
was used to calculate multiplicities at $\sqrt{s} \approx 30$ GeV.
The  $\pi^0$ measurements are done at central rapidities  only.
In order to calculate the rapidity integrated distributions
needed for comparison with the model, we
use the following approximation: 
$dN/dm_T = d^2N/(dm_T dy) \cdot \Delta y$, where $\Delta y = 1$.
Measurements of the transverse momentum spectra of $\pi^0$
mesons performed by several experiments at the ISR \cite{pi30} differ typically
by a factor of about 2. 
This together with similar uncertainty introduced by the
extrapolation procedure for rapidity integrated distributions,
yield a relatively large systematic error on the absolute
normalisation of $\pi^0$ spectra.
An additional bias of a similar magnitude is possible due
to a missing  correction for contributions from resonance decays.
A more detailed analysis is obviously needed in the future.
Nevertheless,
we observe a surprising agreement over more than 10 orders of magnitude
of the experimental results 
with the expectations following from the statistical power--law model.
Typical deviations of the experimental points from the
universal power--law dependence are of about factor 2,
which is a similar magnitude as the systematic errors
estimated above.
In the power--law fits to the data the statistical and
50\% systematic errors were added in squares.
The values of the parameter $P$ ($\cong 10$)
and the noramlization constant $C$ resulting from the
separate fits to the  $m_T$  spectra and hadron
multiplicities are similar (for details see Table 1).
The solid
line in Fig.~1 indicates a fit to the multiplicity data.

As seen from Fig.~1 the power--law model
(\ref{mTp},~\ref{mp}) works reasonably well even for intermediate values
of $m_{(T)}$ ($0.5\div 2$ GeV/c$^2$), where the exponential distributions
(\ref{mTexp},\ref{mexp}) are normally assumed, 
e.g. the power--law function (\ref{mp}) seems to  describe also
the data for $\eta$ and $\phi$ multiplicities. 
This fact requires a further
study.

In order to check whether the observed power--law
$m_T$--scaling is valid also for higher collision
energies  we compiled data on $J/\psi$, $\psi'$
\cite{Ab:97},
$\Upsilon(1s)$, $\Upsilon(2s)$ and $\Upsilon(3s)$
\cite{Ab:95}
in p+$\overline{\rm{p}}$ interactions at 
$\sqrt{s}$ = 1800 GeV
(the highest energy accessible today).
The $p_T$ distributions for the  quarkonia
are measured at midrapidity and they allow to calculate 
$\rho(m_T)$ spectra in this region.
The resulting distributions are indicated by open  symbols
in Fig. 1 together with the line showing
the power--law fit ($P \cong 8$, for details see Table 1) to these data.
It is seen that the distributions of quarkonia 
produced in p+$\overline{\rm{p}}$ interactions at
$\sqrt{s}$ = 1800 GeV follow approximately
the same power--law function.
The relative comparison of  data for different quarkonia may be
biased by a systematic error of about 50\%
due to the use
of midrapidity spectra instead of spectra integrated over all
rapidities. Therefore, as previously, in the fit the square of the error
was calculated as a sum of the squares of statistical and 
systematic (50\%) errors.  

There are no data on $\pi^0$ meson production in
p+$\overline{\rm{p}}$ interactions at $\sqrt{s}$ = 1800 GeV.
However, there are measurements of $\pi^0$ spectra performed
close to midrapidity ($y^* \approx 1.4$) 
at $\sqrt{s}$ = 540 GeV which extend up
to $p_T \cong$ 40 GeV/c \cite{Ba:85}.
The resulting $\rho(m_T)$ distribution is shown in Fig. 2.
The fit of the power--law function indicated by a solid line
yields $P \cong 8$ (for details see Table 1),
the value similar to one obtained for quarkonium spectra at 1800 GeV.
The power--law fits to 30 and 1800 GeV data, indicated in
Fig. 2 for a comparison,
are both below the spectrum at 540 GeV. 
This  observation may suggest a possible violation of the scaling
for $\pi^0$ mesons at 1800 GeV.
This is because
one may expect a monotonic increase of $\rho(m_T)$ at fixed $m_T$ with
increasing collisions energy and this expectation is not obeyed by the 540 GeV data.
We note, however, that the measurements of $\pi^0$ spectrum at high $p_T$ region
were indirect due to difficulty to resolve single photons from $\pi^0$ decay.
Therefore, a direct measurements of  $\pi^0$ spectrum at 1800 GeV are necessary
in order to verify a validity of the scaling in this domain.  

\vspace{0.5cm}

What is the origin of the 
power--law $m_T$--scaling observed in 
p($\overline{\rm{p}}$)+p interactions and 
not (yet) seen in the collisions of heavy nuclei \cite{au}?
The production of heavy hadrons  (like $J/\psi$ 
and $\Upsilon$ mesons)
as well as the power--law behaviour of pion spectra at high $p_T$ 
are fitted by QCD inspired models \cite{parton,Ma:95}.
However different assumptions and 
parameters enter in these models for $\pi^0$ and
quarkonia description.
Thus the observed power--law $m_T$--scaling seems to be
an unexpected and interesting  feature of the data 
which requires explanation. 
In this context we list known to us developments which can be
helpful in attempts to understand the power--law $m_T$--scaling.
It was recently pointed out \cite{Bi:99} that quantum fluctuations of the string
tension may account for the thermal features of the hadron spectra.
A possible appearance of the power--law instead of Boltzmann spectrum
was suggested within thermal field theory \cite{Ma:00} in the low temperature and
high mass limit.
Generalization of standard thermodynamics to non--extensive systems
\cite{Ts:99} leads to distribution of energy in the canonical ensemble
with the power--law tail for high energies.

Finally we 
repeat the basic results  presented in this letter. 
Properly normalized multiplicities and $m_T$ spectra 
of neutral mesons
from $\eta$ to $\Upsilon$ ($m\cong 0.5 \div 10 $~GeV/c$^2$)
and 
$m_T$
spectra of $\pi^0$  meson ($m_T > 1$~GeV/c$^2$)
obey the $m_T$--scaling.
The scaling function has approximately  the power--law form:
$m_{(T)}^{-P}$.
The values of  parameter $P$ and normalisation
constant $C$ resulting from the fits to data on production of different mesons
at fixed collision energy are similar.
This scaling behaviour resembles that expected
in statistical mechanics:
the parameter $P$ plays the role of
temperature 
and the normalisation constant $C$ is analogous to the system volume.
Thus the basic modification of the statistical approach
needed to reproduce the experimental results on hadron production
in p($\overline{\rm{p}}$)+p 
interactions in the large $m_{(T)}$ domain is the change of the
shape of the distribution function.
The Boltzmann function $e^{-E^*/T}$ appearing in the standard 
statistical mechanics has to be substituted by a power--law function
$(E^*/\Lambda)^{-P}$.

\vspace{1cm}
\noindent
{\bf Acknowledgements}

We thank F. Becattini, K. A. Bugaev,  L. Gerland, L.~Frankfurt, 
Sh.~Matsumoto, C. Tsallis,
M.~Thoma, R. Renfordt, W.~Retyk  and M.~Yoshimura for
discussion and comments to the manuscript.
We acknowledge financial support of DAAD, Germany.
The research described in this publication was made possible in part by
Award \# UP1-2119 of the U.S. Civilian Research and the Development
Foundation for the Independent States of the Former Soviet Union
(CRDF).

\newpage
{\bf Table 1}
\noindent
The values of the  parameters ($P$ and $C$) resulting from the
power--law fits ($\rho(m_{(T)})  = C \cdot m_{(T)}^{-P}$) to various
data sets discussed in this letter (for details see text).
The values of $\chi^2/ndf$  are also
given.

\vspace{1.5cm}

\begin{tabular}{|c|c|c|c|}
\hline
          &     &           &                 \cr
   data set    &   $P$      &    $C\cdot10^2$ (GeV$^(P-3)$)   &   $\chi^2/ndf$        \cr
          &     &           &                 \cr
\hline
\hline
          &     &           &                 \cr
~~~neutral meson multiplicity  at $\cong$30 GeV~~~ &~~~ 10.1 $\pm$ 0.3~~~ 
 &~~~ 3.6 $\pm$ 1.0~~~ &~~~ 1.02 ~~~  \cr
          &     &           &                 \cr
 $m_T$--spectra of $\pi^0$ at $\cong$30 GeV & 9.8 $\pm$ 0.1 & 6.1 $\pm$ 0.9  & 1.4          \cr
          &     &           &                 \cr
 $m_T$--spectra of quarkonia at 1800 GeV & 7.7 $\pm$ 0.4 &~~ 3.4~(+4.4,-2.0)~~  & 1.13        \cr
          &     &           &                 \cr
 $m_T$--spectra of $\pi^0$ at 540 GeV & 8.1 $\pm$ 0.1 & 129 $\pm$ 15  & 1.3         \cr
          &     &           &                 \cr
\hline
\end{tabular}

\newpage

\begin{figure}[p]
\epsfig{file=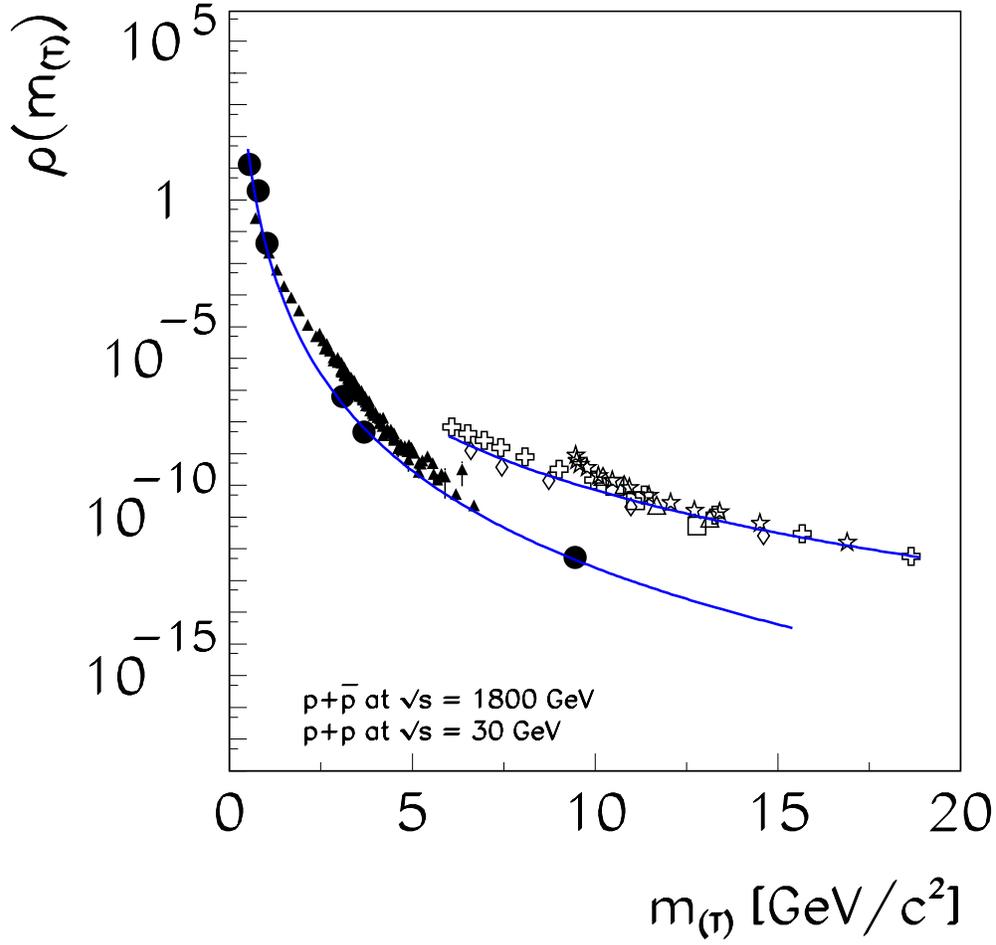,width=14cm}
\vspace{0.5cm}
\caption{
Mean multiplicity of neutral
mesons (full dots) scaled to $\rho(m)$ (\ref{mp}) 
and transverse mass spectra  of $\pi^0$
mesons (full triangles) scaled to $\rho(m_T)$ (\ref{mTp})  
produced in p+p interactions at $\sqrt{s} \approx$~30~GeV 
as well as the quarkonium $\rho(m_T)$ spectra 
($J/\psi$ -- crosses, $\psi'$ -- diamonds,
$\Upsilon(1s)$ -- stars, $\Upsilon(2s)$ -- triangles,
$\Upsilon(3s)$ -- squares)
for p+$\overline{\rm{p}}$ at $\sqrt{s}$ = 1800 GeV.
The solid lines indicate power--law fits to the  data.
}
\label{fig1}
\end{figure}

\newpage

\begin{figure}[p]
\epsfig{file=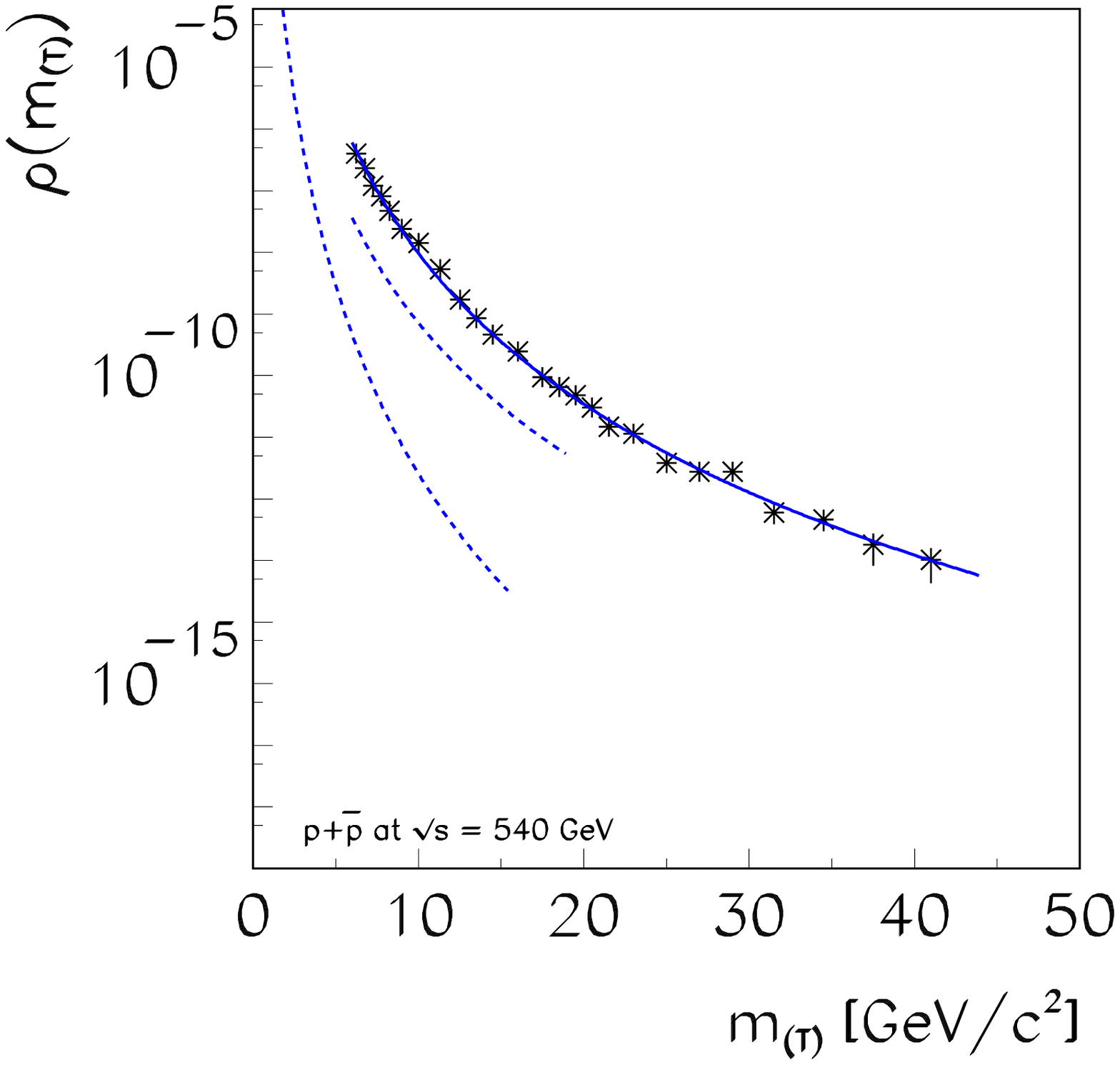,width=14cm}
\vspace{0.5cm}
\caption{The  $\rho(m_T)$ spectra of $\pi^0$ meson
for p+$\overline{\rm{p}}$ interactions at
$\sqrt{s}$ = 540 GeV.
The solid line indicates a power--law fit to the  540 GeV data,
whereas the two dashed lines show the corresponding  fits to
the results at 30 and 1800 GeV (see Fig. 1).
}
\label{fig2}
\end{figure}


\newpage


\begin{thebibliography}{99}




\bibitem{Ha:65}
R. Hagedorn, Suppl. Nuovo Cimento {\bf 3} (1965) 147. 

\bibitem{Be:97}
F. Becattini and U. Heinz, Z. Phys. {\bf C76} (1997) 269,\\
F. Becattini, Nucl. Phys. Proc. Suppl. {\bf 92} (2001) 137, \\  
P. V. Chliapnikov, Phys. Lett. {\bf B462} (1999) 341 and 
{\bf B470} (1999) 263. 

\bibitem{Da:80}
P. Darriulat, Ann. Rev. Nucl. Part. Sci. 
{\bf 30} (1980) 159 and references therein.

\bibitem{parton}
R. D. Field and R. P. Feynman, 
Phys. Rev. {\bf D15} (1977) 2590, \\
R. P. Feynman, R. D. Field and G. C. Fox,
Phys. Rev. {\bf D18} (1978) 3320, \\
F. M. Borzumati and G. Kramer,
Z. Phys. {\bf C67} (1995) 137.

\bibitem{Ga:99}
M. Ga\'zdzicki and M. I. Gorenstein,
Phys. Rev. Lett. {\bf 83} (1999) 4009.

\bibitem{Al:96}
T. Alexopoulos et al. (E771 Collab.), 
Phys. Lett. {\bf B374} (1996) 271.

\bibitem{Ra:80}
J. Rafelski and M. Danos, 
Phys. Lett. {\bf B97} (1980) 279.

\bibitem{Ag:91}
M. Aguilar--Benitez et al. (LEBC-EHS Collab.), 
Z. Phys. {\bf C50} (1991) 405.

\bibitem{Ko:80}
C. Kourkoumelis et al.,
Phys. Lett. {\bf 91B} (1980) 481.

\bibitem{pi30}
K. Eggert et al., Nucl. Phys. {\bf B98} (1975) 49, \\
A. Angelis et al., Phys. Lett. {\bf 79B} (1978) 505, \\
F. Busser et al., Phys. Lett. {\bf 46B} (1973) 471 and
Nucl. Phys. {\bf B106} (1976) 1, \\
C. Kourkoumelis et al., Z. Phys. {\bf C5} (1980) 95.

\bibitem{Ab:97}
F. Abe et al. (CDF Collab.), 
Phys. Rev. Lett. {\bf 79} (1997) 572.

\bibitem{Ab:95}
F. Abe et al. (CDF Collab.), 
Phys. Rev. Lett. {\bf 75} (1995) 4358.

\bibitem{Ba:85}
M. Banner et al. (UA2 Collab.), 
Z. Phys. {\bf C27} (1985) 329.

\bibitem{au}
J. Harris et al. (STAR Collab.), Preceedings of Quark Matter
Conference, Stony Brook, January 2001.

\bibitem{Ma:95}
M.  L. Mangano, hep-ph/9507353,\\
G. A. Schuler, Z. Phys. {\bf C71} (1996) 317,\\
R. Vogt, Phys. Rep. {\bf 310} (1999) 197.

\bibitem{Bi:99}
A. Bia{\l}as, Phys. Lett. {\bf B466} (1999) 301.

\bibitem{Ma:00}
Sh. Matsumoto and M. Yoshimura,
Phys. Rev. {\bf D61}, 123508 (2000).

\bibitem{Ts:99}
C. Tsallis, J. Stat. Phys. {\bf 52}, 479 (1988),\\
for recent review see also
Braz. J. Phys. {\bf 29} (1999) 1.

\end{thebibliography}
\end{document}